# A proposal to systematize introducing DevOps into the software development process

Luciano de Aguiar Monteiro
Center of Advanced Studies and Systems of Recife
Teresina, Brazil
lam@cesar.school

*Abstract*— The software development industry has been evolving with new development standards and service delivery models. Agile methodologies have reached their completion with DevOps, thereby increasing the quality of the software and creating greater speed in delivery. However, a gap regarding the formalization of its adoption and implementation doubts became relevant. My hypothesis is that, by systematizing the introduction of DevOps into the software development process and defining the function of the members of the DevOps team members, may well make it quicker to implement this process, thus reducing conflicts between the teams. As part of the investigation of this hypothesis, the result of the research will be applied in practical development environments i.e. in a Technology Agency of the State of the Brazilian Government and also at the Brazilian Company Neurotech in order to evaluate its effectiveness from metrics appropriate for DevOps environments.

## I. Problem

DevOps is an emerging practice that has been adopted in the software development cycle. It focuses on the convergence of standards between the **Dev**elopment teams and the **Op**erations teams and it seeks to improve cooperation between both teams, hence the origin of the term [1].

However, there is no consensus on the definition of what DevOps is. Wiedemann et al. [2] emphasize that one of the biggest challenges in the industry is the lack of a formal concept for DevOps. This conceptual flaw in what DevOps is, directly impacts the understanding of the objects and actions needed to overcome this flaw [3].

The adoption of Agile and DevOps Methodologies continues to grow, driven by the "need for speed", agility and flexibility, evidenced in the World Quality Report of 2019 [4], in which 99% of the interviewees said they were using DevOps in at least some of their business.

Implementing DevOps has become more difficult due to the lack of formalizing the concept and adoption processes. Although there is an abundance of information, practices and tools related to DevOps, it is still unclear how anyone could take advantage of this rich, yet diffuse information in an organized and structured way to properly adopt DevOps [6].

The study by Zulfahmi [7] provides evidence of this lack of standardization. The author also states that other major challenges are the lack of process and guidelines for implementing the practice of DevOps in Continuous Delivery.

In the Systematic Literature Review conducted by Gasparaite; Naudziunaite and Ragaisis [8], 24 DevOps models were identified, but only 04 were considered applicable for practical use: the Focus Area, Bucena-Kirikova, Mohamed, and Radstaak models. However, in the Radstaak model, not all the steps necessary for its practical use have been described; the Mohamed model did not present the evaluation process, only how to apply the model; and the Focus Area and Bucena-Kirikova Models can be applied in practice, since their authors document the evaluation methods adopted in academic publications.

Another aspect to be highlighted concerns what must be done or would be appropriate prior to an organization adopting DevOps. Leite et al. [9] ponder this aspect by considering a difficult question that the literature has not yet fully answered. They further point out that it is incomplete and even contradictory in relation to the subject.

At the same time, the variety of DevOps tools seems to challenge the idea of there being a single person with the role of administering the entire process. Even mature teams, who have both knowledge of development and infrastructure operations, may find it difficult to be familiar with all these tools, thus making it necessary to define the roles of those involved in this process.

The need to systematize the introduction of DevOps in the software development process is therefore necessary. Thus, the need for more research and empirical work is essential to put into practice and validate a proposal to systematize the introduction of DevOps.

In view of the above, the following problems are in evidence: a) the lack of a conceptual definition and, consequently, the need for a proposal to systematize the introduction of DevOps in the software development process, that presents systematic improvements, evidenced by its adoption and the results of its effectiveness; and b) the lack of definition of the roles of those involved in this cycle.

The present thesis should conduct a study that sets out to identify the sets of good practices already used in software development processes that use DevOps, and based on these to conduct an analysis, validation and standardization of their use, with a view to systematizing the introduction and execution of DevOps in this process.

## II. Research Hypothesis

A software production line that uses DevOps, has a well-defined automation cycle, starting with the developers' source code commits for the code version-control system. When the CI server identifies the completion of the commits, it performs the necessary tests and, if necessary, provides feedback to the developers [10]. In this thesis we will propose systematic practices for introducing DevOps and improving the efficiency of the software production process that uses DevOps.

As a way of conducting the research and delimiting the scope of the study, the following hypothesis is proposed:

> Having adopted a set of systematized practices for introducing DevOps, and for formalizing and delimiting the role of the members of the DevOps team in the Software Production Process, Software Factories become faster in their implementation, thereby reducing conflicts between teams and providing quality deliverables.

To validate the hypothesis, the following questions were constructed:

**RQ1**: What are the gaps in the software development process that use DevOps practices?

In order to propose solutions to the existing problems when introducing DevOps, one must first identify the gaps in the process. The purpose of RQ1 is to locate these gaps and the actions that must be taken to resolve such issues.

**RQ2**: What are the practices that support the systematic introduction of DevOps in the software development process?

The objective of RQ2 is to identify all the good practices used by the development and academic industry that support introducing it into the software development process.

### III. EXPECTED CONTRIBUTION

The objective of my doctoral thesis is to improve the software development process using DevOps and to propose new ways of introducing it. The expected contributions of this project are summarized below:

1) to systematize the introduction of DevOps in the software development process; and
2) to improve the adoption of DevOps in the software development process.

### IV. METHODS AND PRELIMINARY FINDINGS

As a preliminary study, a Systematic Mapping (SM) of Literature [11] was undertaken, in May 2020, with a view to analyzing the topic, and hence seeking to identify existing gaps in the area of DevOps and difficulties in the implementation process.

The SM [11] will serve to underpin the initial conduct of the research, and thus provide a diagnosis of the software production process that uses DevOps, also providing initial data that can be used to build a set of systematic practices for adopting DevOps.

The SM [11] showed how DevOps is highlighting the following issues: pipeline security, effective adoption of a cloud environment, adoption of microservices, infrastructure as code, use of container solutions, and tools to automate the development pipeline. Regarding best practices, DevOps was found to have: Infrastructure as Code, Continuous Integration, Continuous Delivery, Continuous Deployment.

One of the questions raised in the MS [11] was about the existence of Maturity Models, in which 11 were found. However, only two were validated by companies, and one of them is applied only in IBM solutions. The roles of the actors involved in the process were found in 02 studies, with different names, but performing the same function.

In a second phase, a Multi-Vocal Literature Review will be carried out to look for relevant information on emerging industry topics, which were not achieved by using SM. This type of Review has both academic publications and gray literature as input, its main objective being to close the gap between academic research and professional practice [12].

The methodology adopted for practical research will be Design Science Research (DSR), which is a method that establishes and operationalizes research when the desired objective is an artifact or a recommendation [13]. Given the above, there will be a need for an iterative cycle and the production of an artifact, which may be a set of systematic practices for adopting DevOps, which will be validated and improved throughout the research.

The research will be applied in a Technology Agency of the State of the Brazilian Government, of which the author is Technical Director and has an Information Systems Coordination Unit with several applications under development using DevOps practices, and in the Brazilian company Neurotech, which develops advanced solutions of Artificial Intelligence, Machine Learning and Big Data, and has a portfolio of more than 100 customers.

### V. EVALUATION PROCEDURES

I am starting my second year of doctorate and by its end, RQ1 and RQ2 will have been answered. The next step will be to propose a handbook that will systematize the introduction of DevOps, during which it will be applied in practice as previously mentioned in a State Technology Agency and Neurotech.

The instrument for evaluating the results of the research proposed in the DevOps development process will be conducted by means of specific DevOps metrics, collected throughout the development process. Some studies have presented a set of metrics that can be used to assess the applicability of DevOps in the Software Production Process: Delivery time; Distribution frequency; the Percentage of changes that fail; Recovery time (from a failure in the production environment); a culture of capturing and using measures across the organization; a repository at the organizational level to store measurement data; a set of KPIs to assess performance; a standardized method for reporting raw / analyzed metrics; and, quality of code metrics (defects, security vulnerabilities, technical debt) [14][15][16][17][18].

It is expected that, by using the data collection methods presented, and assessing the systematic efficiency of introducing the proposed DevOps into the development process, this study will contribute to the academic community, as well as to the systems development industry in such a way as to improve the efficiency of the software development process when adopting DevOps, by having an innovative methodology for its introduction.